\documentclass[twocolumn,showpacs,pra,aps,graphicx,preprintnumbers,amsmath,amssymb]{revtex4}

\usepackage{dcolumn}
\usepackage{bm}

\usepackage{epic}
\usepackage{eepic}
\usepackage{graphicx}

\newcommand{\ket}[1]{|#1\rangle}

\newcommand{\Order}[1]{\mathcal{O}(#1)}

\begin{document}

\title{Number statistics of molecules formed from ultra-cold atoms }
\author{D. Meiser}
\author{P. Meystre}
\affiliation{Optical Sciences Center, The University of Arizona,
Tucson, AZ 85721}


\begin{abstract}
We calculate the number statistics of a single-mode molecular
field excited by photoassociation or via a Feshbach resonance from
an atomic Bose-Einstein condensate (BEC), a normal atomic Fermi
gas and a Fermi system with pair correlations (BCS state). We find
that the molecule formation from a BEC is a collective process
that leads for short times to a coherent molecular state in the
quantum optical sense. Atoms in a normal Fermi gas, on the other
hand, are converted into molecules independently of each other and
result for short times in a molecular state analogous to that of a
classical chaotic light source. The BCS situation is intermediate
between the two and goes from producing an incoherent to a
coherent molecular field with increasing gap parameter.
\end{abstract}

\maketitle

A remarkable development following the availability of ultracold,
quantum-degenerate atomic samples has been the coherent generation
of molecular dimers by means of Feshbach resonances
\cite{Inouye:MoleculeFR,Donley:MoleculeBEC,Duerr:MoleculeRP} or
via two-photon Raman transitions \cite{Wynar:MoleculePa} from
atomic Bose-Einstein condensates (BEC). These trailblazing
experiments were soon followed by the generation of molecular
dimers from two fermionic atoms
\cite{Greiner:MoleculeBEC,Zwierlein:Li2,Jochim:Li2}, in which case
the capability to tune the strength and change the sign of the two-body
interaction between fermions opens up the possibility to carry out
detailed studies of many longstanding questions concerned with the
so-called BEC-BCS crossover
\cite{Timmermans:BCS,Regal:BEC_BCScrossover,Bartenstein:BEC_BCScrossover,Zwierlein:BEC_BCScrossover}.
The observation of Feshbach resonances between atoms of different
species \cite{Stan:heteromolecules,Inouye:heteromolecules} also
opens up the possibility of generating quantum-degenerate gases of
heteronuclear molecules. These experiments have triggered a number
of theoretical studies aimed at a better understanding of the
process of coherent molecule formation and the properties of the
coupled atom-molecule system
\cite{Duine:Microscopic,Holland:BCS,Ohashi:BCSBECcrossover}, but
the description of the dynamics of these systems remains
incomplete. In particular, the dependence of the quantum
statistical properties of the resulting molecular field on those
of the initial atoms is largely unknown.

In this letter we apply methods of quantum optics to the coupled
atom-molecule system and analyze the number statistics of the
molecules formed when starting from an atomic BEC, a normal Fermi
gas, and a Fermi system with pairing (BCS type state). We restrict
ourselves to zero temperature $T=0$ throughout, and further assume
that the molecules are formed in a single mode, typically the
ground state of a trap. The statistical properties of the
molecular field are then determined by numerically integrating the
relevant Schr\"odinger equations, and further analytical insight
is provided by time-dependent perturbation theory considerations.
We find that the statistics of the molecular field provide a
distinct signature of the initial atomic state, thereby offering a
powerful diagnostic tool to measure atomic statistical properties
such as e.g. the presence of BCS pairing.

Consider first a cloud of weakly interacting bosons well below the
condensation temperature $T_c$. It is a good approximation to
assume that all atoms are in the condensate, described by the
condensate wave function $\chi_0(x)$ with energy eigenvalue $\mu$.

Atom pairs can be transformed into trapped molecules by means of
photo-association or by means of a Feshbach-resonance.  The coupled system of
atoms and molecules is described by the effective two-mode Hamiltonian
\cite{Javanainen:Twomodemodel,Anglin:twomode}
\begin{equation}
\hat{H}_{\text{BEC}}=\delta \hat{a}^\dagger \hat{a} +
g\left(\hat{a}^\dagger \hat{c}^2 + \hat{a} \hat{c}^{\dagger
2}\right). \label{twomode_hamiltonian}
\end{equation}
where $\hat{a}^\dagger$ and $\hat{c}$, $\hat{c}^\dagger$ are the bosonic
annihilation and creation operators for the atoms in the condensate and for the
molecules, respectively, $\delta$ is the detuning between the molecular and
atomic level, $g$ is the effective coupling constant, and we set $\hbar = 1$
throughout.

Typical experiments start out with all the particles in the
condensate and no molecules, corresponding to the initial state
\begin{equation}
\ket{\psi(t=0)}=\frac{\hat{c}^{\dagger{N_a}}}{\sqrt{N_a!}}\ket{0}.
\label{BEC_initial}
\end{equation}
where $N_a=2N_{\rm max}$ is the number of atoms, $N_{\rm max}$ is
the maximum possible number of molecules and $\ket{0}$ is the
vacuum of both molecules and atoms. For simplicity we only
consider the case of even numbers of atoms, but our treatment can
easily be extended to odd atom numbers. The Hamiltonian
\eqref{twomode_hamiltonian} clearly conserves the total number of
free and bound atoms, $2\hat{a}^\dagger \hat{a}+\hat{c}^\dagger
\hat{c}$. Therefore, the evolution of the system from the initial
state \eqref{BEC_initial} can be described in the basis
\begin{equation}
\ket{\phi_n}=\frac{\hat{a}^{\dagger n} {\hat
c}^{2n}\ket{\psi(t=0)}} {\mathcal N},\quad n=0,1,\ldots,N_{\rm
max},
\end{equation}
with $\mathcal N$ being a normalization constant. The state
$\ket{\phi_n}$ corresponds to $n$ molecules and $2(N_{\rm
max}-n)$ atoms.

Expanding the state of the system on this basis as
$\ket{\psi(t)}=\sum_n y_n \ket{\phi_n}$ we can write the Schr\"odinger
equation as
\begin{multline}
i\frac{dy_n}{dt}=\sqrt{N-n+1}\sqrt{N-n+2}\sqrt{n}y_{n-1}\\+
\sqrt{N-n}\sqrt{N-n-1}\sqrt{n+1}y_{n+1}+\delta n y_n.
\end{multline}
Solving this coupled set of $N_{\rm max}$ equations numerically
gives then the probability $P_n(t)=|y_n(t)|^2$ of finding $n$
molecules in the single-mode molecular field, the ``molecule
statistics.''

\begin{figure}
\includegraphics[width=0.9\columnwidth]{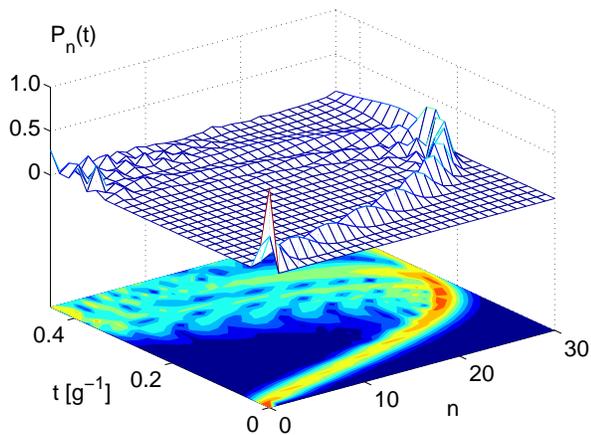}
\caption{Number statistics of molecules formed from a BEC with $N_{\rm max}=30$
and $\delta=0$.}
\label{PnBEC}
\end{figure}

Figure \ref{PnBEC} shows $P_n(t)$ for 30 initial atom pairs and
$\delta=0$. Starting in the state with zero molecules, a
wave-packet-like structure forms and propagates in the direction
of increasing $n$. Near $N_{\rm max}$ the molecules begin to
dissociate back into atom pairs. (Note that our two-mode model
neglects ``rogue photodissociation,'' an approximation appropriate
for short enough times so that the condensate is not significantly
depleted \cite{Javanainen:Twomodemodel}.)

We can gain further insight into the short-time dynamics of
molecule formation by using first-order perturbation theory, which
gives the mean molecule number
\begin{equation}
n(t)=\langle \hat{a}^\dagger \hat{a}\rangle= (gt)^2 2N_{\rm
max}(2N_{\rm max}-1) \label{nBEC}
\end{equation}
and the second factorial moment
\begin{eqnarray}
g^{(2)}(t_1,t_2)&\equiv&
\frac{\langle \hat{a}^\dagger(t_1)\hat{a}^\dagger (t_2)
\hat{a}(t_2)\hat{a}(t_1)\rangle}{\langle\hat{a}^\dagger(t_1)
\hat{a}(t_1)\rangle\langle\hat{a}^\dagger(t_2)\hat{a}(t_2)\rangle}\\
&=&\frac{(2N_{\rm max}-2)(2N_{\rm max}-3)}{2N_{\rm max}(2N_{\rm max}-1)}\nonumber\\
&=&1-\frac{2}{N_{\rm max}}+\Order{N_{\rm max}^{-2}}\nonumber.
\end{eqnarray}
For $N_{\rm max}$ large enough we have $g^{(2)}(t_1,t_2)
\rightarrow 1$, the value characteristic of a Glauber coherent
field. From $g^{(2)}$ and $n(t)$ we also find the relative width
of the molecule number distribution as
\begin{equation}
\frac{\sqrt{\langle(\hat{n}-n)^2\rangle}}{n}= \sqrt{g^{(2)} +
n^{-1}-1}. \label{deltanBEC}
\end{equation}
It approaches $n^{-1/2}$ in the limit of large $N_{\rm max}$,
typical of a Poisson distribution. This confirms that for short
enough times, the molecular field is coherent in the sense of
quantum optics.

Let us now turn to the case of photoassociation from two different species of
non-interacting ultra-cold fermions. The two species are denoted by spin up and
down. At $T=0$, the atoms fill a Fermi sea up to an energy $\mu$. Weak
repulsive interactions give rise only to minor quantitative modifications that
we ignore in this paper. We refer to this system of non-interacting Fermions as
a normal Fermi gas (NFG) \cite{LandauIX}.

As before we assume that atom pairs are coupled only to a single
mode of the molecular field which we assume to have zero momentum
for simplicity. Only pairs of atoms with opposite momenta $+k$ and
$-k$ and opposite spin can then be combined into molecules.
Introducing pseudo-spin operators in the subspace of each pair of atoms with
opposite momenta and spin as
\begin{equation}
\hat{\sigma}_{(k)}^z=\hat{c}_{k,\uparrow}^\dagger \hat{c}_{k,\uparrow}+
\hat{c}_{-k,\downarrow}^\dagger \hat{c}_{-k,\downarrow} - 1,
\quad \hat{\sigma}_{(k)}^+
= \hat{\sigma}_{(k)}^{-\dagger} =\hat{c}_{k,\uparrow}^\dagger
\hat{c}_{-k,\downarrow}^\dagger,
\end{equation}
where $\hat{c}_{k,\uparrow\downarrow}$ and
$\hat{c}_{k,\uparrow\downarrow}^\dagger$ are the annihilation and
creation operators for a fermion of momentum $k$ with up and down spin, the
Hamiltonian describing the coupled atom-molecule system takes the form
\cite{Javanainen,Barankov}
\begin{equation}
\hat{H}_{\text{NFG}}= \sum_k E_k \hat{\sigma}_{(k)}^z + \delta
\hat{a}^\dagger \hat{a} + g\left ( \hat{a}^\dagger\sum_k
\hat{\sigma}_{(k)}^- + H.c. \right ), \label{fermions_hamiltonian}
\end{equation}
where $E_k=k^2/2M$ is the kinetic energy of an atom of momentum
$k$ and mass $M$ and $\delta$ is the detuning between the molecule
energy and the Fermi energy. This Hamiltonian is formally
equivalent to the well studied Tavis-Cummings model of quantum
optics, which describes the coupling of an ensemble of two-level
atoms to a single mode of the electromagnetic field
\cite{Tavis:Tavis_Cummings}.

The atomic kinetic energies $E_k$, which in this analogy
correspond to the energies of the fictitious two-level atoms, give
rise to inhomogeneous broadening. In case this broadening can be
neglected the atoms are conveniently described in terms of the
eigenstates $\ket{S,m}$ of $\hat{S}^{2}$ and $\hat{S}^z$,
\begin{equation}
\hat{S}^2\ket{S,m}=S(S+1)\ket{S,m},\quad \hat{S}^z\ket{S,m}=m\ket{S,m},
\end{equation}
where $\hat{S}^l=\sum \hat{\sigma}_{(k)}^l,\quad l=x,y,z$, and we
have as usual $\hat{\sigma}_{(k)}^{x}=(\hat{\sigma}_{(k)}^+ +
\hat{\sigma}_{(k)}^-)/2$ and
$\hat{\sigma}_{(k)}^{y}=(\hat{\sigma}_{(k)}^+ -
\hat{\sigma}_{(k)}^-)/2i$. In that limit the Hamiltonian
(\ref{fermions_hamiltonian}) commutes with
$\hat{S}^z+\hat{a}^\dagger \hat{a}$ and, like in the BEC case, the
system can be uniquely described by the probability amplitudes
$y_n$ for finding $n$ molecules. This degenerate model is studied
in detail in Ref. \cite{Miyakawa:degeneratemodel}. We focus here
instead on the situation where the kinetic energy of the atoms
cannot be neglected.

\begin{figure}
\includegraphics[width=0.9\columnwidth]{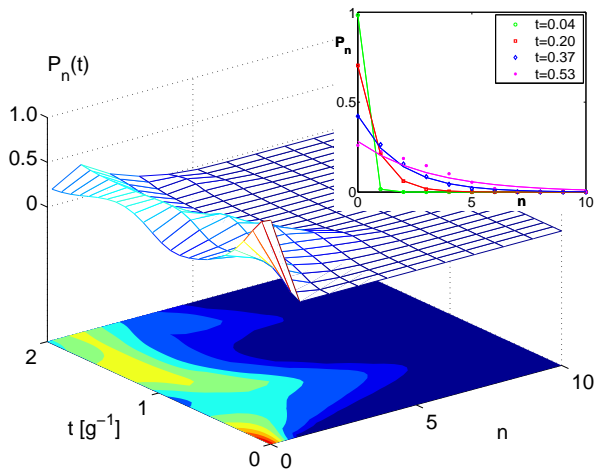}
\caption{Number statistics of molecules formed from a
normal Fermi gas. This simulation is for $N_a=20$ atoms,
the detuning is $\delta=0$ and the Fermi energy is $\mu=0.1g$.
The inset shows fits of the number statistics to thermal
distributions for various times.}
\label{PnNFG}
\end{figure}

Figure \ref{PnNFG} shows the molecule statistics obtained by a
numerical integration of the Schr\"odinger equation corresponding
to the Hamiltonian \eqref{fermions_hamiltonian}. The result is
clearly both qualitatively and quantitatively very different from
the case of molecule formation from an atomic BEC.

From the Tavis-Cummings model analogy we expect that for short
times the statistics of the molecular field should be chaotic, or
``thermal'', much like those of a single-mode chaotic light field.
This is because each individual atom pair ''emits'' a molecule
independently and without any phase relation with other pairs.
That this is the case is illustrated in the inset of Fig.
\ref{PnNFG}, which fits the molecule statistics at selected short
times with chaotic distributions of the form
\begin{equation}
P_{n,\text{thermal}}=\frac{e^{-n/\langle n\rangle}}{\sum_n
e^{-n/\langle n \rangle}} .\label{thermaldistribution_eqn}
\end{equation}
The increasing `pseudo-temperature' $\langle n \rangle$
corresponds to the growing average number of molecules as a
function of time.

We can again determine the short-time properties of the molecular
field in first-order perturbation theory. For the mean number of
molecules we find
\begin{equation}
n(t)=(gt)^2 2N_{a}.
\end{equation}
It is proportional to $N_a$, in contrast to the BEC result, where
$n$ was proportional to $N_a^2$, see Eq.~(\ref{nBEC}). This is
another manifestation of the independence of all the atom pairs
from each other: While in the BEC case the molecule production is
a collective effect with contributions from all possible atom
pairs adding constructively, there is no such collective
enhancement in the case of Fermions. Each atom can pair up with
only one other atom to form a molecule.

For the second factorial moment we find
\begin{equation}
g^{(2)}(t_1,t_2)=2\left(1-\frac{1}{2N_a}\right)
\label{g2NFG}
\end{equation}
which is close to two, typical of a chaotic or thermal field.

Unlike repulsive interactions, attractive interactions between
fermions have a profound impact on molecule formation. It is known
that such interactions give rise to a Cooper instability that
leads to pairing and drastically changes the qualitative
properties of the atomic system. The BCS reduced Hamiltonian
including interactions between the atoms is \cite{Kittel}
\begin{equation}
\hat{H}_{\text{BCS}}=
\hat{H}_{\text{NFG}}-V\sum_{k,k^\prime}
\hat{\sigma}_k^+ \hat{\sigma}_{k^\prime}^-.
\label{BCS_hamiltonian}
\end{equation}
The BCS ground state is found by minimizing $\langle
\hat{H}_\text{BCS}-\mu \hat{N}\rangle$ using the ansatz
\begin{equation}
\ket{\text{BCS}}=\prod_k (u_k+v_k \hat{\sigma}_k^+)\ket{},
\end{equation}
with the result
\begin{gather}
u_k^2=\frac{1}{2}\left(1-\frac{2\xi_k}{\sqrt{\Delta^2+4\xi_k^2}}\right),\\
v_k^2=\frac{1}{2}\left(1+\frac{2\xi_k}{\sqrt{\Delta^2+4\xi_k^2}}\right),
\end{gather}
where $\xi_k=(E_k-\mu)$ is the kinetic energy of the atoms
measured from the Fermi surface and $\Delta = V\sum_k u_k v_k$ is
the gap parameter. It is determined by numerically solving the gap
equation
\begin{equation}
\frac{2}{V}=\sum_k\frac{1}{\sqrt{\Delta^2 + 4\xi_k^2}},
\end{equation}
which is readily done for the small atom numbers at hand. The
dynamics is then obtained by a numerical integration of the
Schr\"odinger equation with $\ket{\text{BCS}}$ as the initial
atomic state and the molecular field in the vacuum state.

\begin{figure}
\includegraphics[width=0.9\columnwidth]{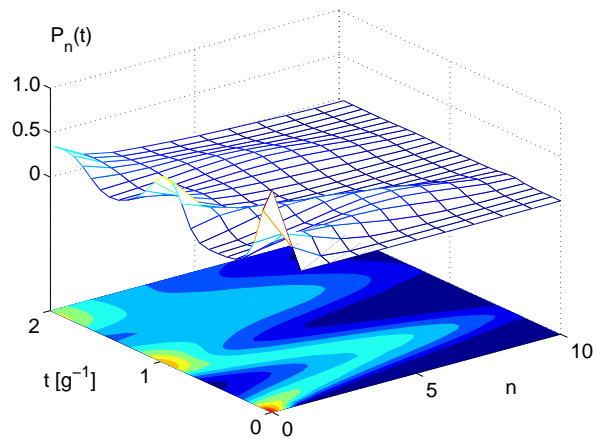}
\caption{Number statistics of molecules formed from a Fermi gas
with pairing correlations. For this simulation the detuning is
$\delta=0$, the Fermi energy is $\mu=0.1g$ and the background
scattering strength is $V=0.03g$ resulting in $N_a\approx 9.4$
atoms and a gap of $\Delta\approx0.15g$.} \label{PnBCS}
\end{figure}

Figure \ref{PnBCS} shows the resulting molecule statistics for
$V=0.03g$, which corresponds to $\Delta=0.15g$ for the system at
hand. (This large background scattering strength was chosen in
order for the gap equation to have a positive solution for the
small particle numbers to which we are limited by computer memory
requirements.) Clearly, the molecule production is much more
effective than in the case of a normal Fermi gas. The molecules
are produced at a higher rate and the maximum number of molecules
is larger. The evolution of the number statistics is reminiscent
of the BEC case.

\begin{figure}
\includegraphics[width=0.8\columnwidth]{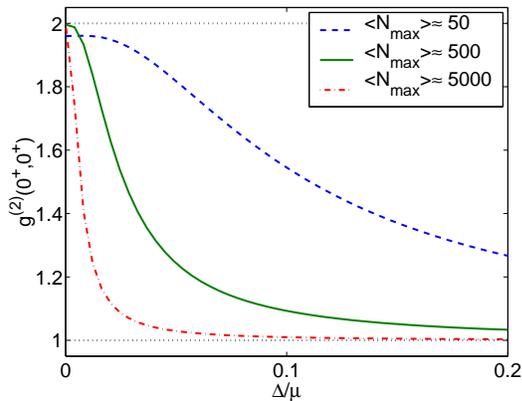}
\caption{$g^{(2)}(0^+,0^+)$ as a function of the gap parameter
$\Delta$.} \label{g2ofgap}
\end{figure}

The short-time dynamics is again obtained in first-order
perturbation theory, which gives now
\begin{eqnarray}
n(t)&=&(gt)^2\left(\sum_{k\neq k^\prime} u_k v_k u_{k^\prime}v_{k^\prime}
+\sum_k v_k^2\right)\\
&\approx&(gt)^2\left[\left(\frac{\Delta}{V}\right)^2 + N_a\right].
\end{eqnarray}
In addition to the term proportional to $N_a$ representing the
incoherent contribution from the individual atom pairs that was
already present in the normal Fermi gas, there is now an
additional contribution proportional to $(\Delta/V)^2$. Since
$(\Delta/V)$ can be interpreted as the number of Cooper pairs in
the quantum-degenerate Fermi gas, this term therefore can be
understood as resulting from the {\em coherent} conversion of
Cooper pairs into molecules in a collective fashion similar to the
BEC case. The coherent contribution results naturally from the
nonlinear coupling of the atomic field to the molecular field.
This nonlinear coupling links higher-order correlations of the
molecular field to lower-order correlations of the atomic field.
For the parameters of Fig. \ref{PnBCS} $\Delta/V\approx 6.5$ so
that the coherent contribution from the Cooper pairs clearly
dominates over the incoherent contribution from the unpaired
fermions. Note that no signature of that term can be found in the
momentum distribution of the atoms themselves. Their momentum
distribution is given by $\langle \hat{c}_{k,\sigma}^\dagger
\hat{c}_{k,\sigma}\rangle = v_k^2, \sigma = \uparrow,\downarrow$ and is very
similar to that of a normal Fermi gas. The
short-time value of $g^{(2)}(t_1,t_2)$, shown in Fig.
\ref{g2ofgap}, decreases from the value \eqref{g2NFG} for a normal
Fermi gas at $\Delta=0$ down to one as $\Delta$ increases,
underlining the transition from incoherent to coherent molecule
production.

In summary, we have applied concepts of quantum optical coherence
to characterize the coherent generation of a molecular field
created from a quantum-degenerate atomic sample. For atoms
initially in a BEC the resulting molecular field is to a good
approximation coherent. This is in sharp contrast to the case of
atoms in a normal Fermi gas, where we have made use of an analogy
with the Tavis-Cummings model to show that the statistics of the
resulting molecular field is very similar to those of a
single-mode chaotic light field. The BCS case interpolates between
the two extremes, with an `incoherent' contribution from unpaired
atoms superposed to a `coherent' contribution from atomic Cooper
pairs. We see, then, that the quantum statistics of the atomic
sample has a profound impact on the quantum statistical properties
of the resulting molecular field. Conversely, these statistics
provide a distinct signature of the initial atomic state and
suggest the use of single molecule counting as a diagnostic tool
for atomic states.

This work is supported in part by the US Office of Naval Research,
by the National Science Foundation, by the US Army Research
Office, and by the National Aeronautics and Space Administration.

\bibliography{draft4}
\end{document}